\documentclass[journal,comsoc]{IEEEtran}
\usepackage[T1]{fontenc}% optional T1 font encoding

\ifCLASSINFOpdf
\else
\fi
\usepackage{graphicx}                
\usepackage{amsmath,mathtools} 
\usepackage{bm}      
\usepackage{cite} 
\usepackage{hyperref}
\usepackage{cleveref}
\makeatletter
\let\MYcaption\@makecaption
\makeatother
\usepackage[switch]{lineno}
\usepackage[font=footnotesize]{subcaption}
\usepackage{xcolor}

\makeatletter
\let\@makecaption\MYcaption
\makeatother
\usepackage[font=footnotesize,labelsep=period]{caption}
\newcommand\numberthis{\addtocounter{equation}{1}\tag{\theequation}}

\newtheorem{theorem}{\textbf{Theorem}}[section]

\DeclareMathOperator{\Tr}{Tr}
\DeclareMathOperator{\rank}{rank}                    
\usepackage[cmintegrals]{newtxmath}
\begin{document}
%
% paper title
% Titles are generally capitalized except for words such as a, an, and, as,
% at, but, by, for, in, nor, of, on, or, the, to and up, which are usually
% not capitalized unless they are the first or last word of the title.
% Linebreaks \\ can be used within to get better formatting as desired.
% Do not put math or special symbols in the title.
\title{Lower Bound on the Sum-rate of Decremental Beam Selection Algorithm for Beamspace MIMO Systems}
%\title{XXX of the YYY ZZZ Algorithm for AAAA Systems}
%
% author names and IEEE memberships
% note positions of commas and nonbreaking spaces ( ~ ) LaTeX will not break
% a structure at a ~ so this keeps an author's name from being broken across
% two lines.
% use \thanks{} to gain access to the first footnote area
% a separate \thanks must be used for each paragraph as LaTeX2e's \thanks
% was not built to handle multiple paragraphs
%
%\iffalse
\author{Naveed Iqbal, \textit{Student Member, IEEE}, Waqas Ahmad, Christian Schneider
	and~Reiner S. Thom\"{a}, \textit{Fellow, IEEE}% <-this % stops a space
	\thanks{Authors N. Iqbal, C. Schneider and R.S. Thom\"{a} are from Technische Universit\"{a}t Ilmenau, Germany. W. Ahmad is with department of computer science, Otago university, Newzealand (e-mail: naveed.iqbal.q@gmail.com).}% <-this % stops a space
	%\thanks{C. Schneider is a senior research assistant at the Electronic Measurement Research Lab  and Prof. R. S. Thom\"{a} is the head of the Electronic Measurement Research Lab at the Technische Universit\"{a}t Ilmenau(e-mail: christian.schneider@tu-ilmenau.de, reiner.thomae@tu-ilmenau.de).}% <-this % stops a space
}

% note the % following the last \IEEEmembership and also \thanks - 
% these prevent an unwanted space from occurring between the last author name
% and the end of the author line. i.e., if you had this:
% 
% \author{....lastname \thanks{...} \thanks{...} }
%                     ^------------^------------^----Do not want these spaces!
%
% a space would be appended to the last name and could cause every name on that
% line to be shifted left slightly. This is one of those "LaTeX things". For
% instance, "\textbf{A} \textbf{B}" will typeset as "A B" not "AB". To get
% "AB" then you have to do: "\textbf{A}\textbf{B}"
% \thanks is no different in this regard, so shield the last } of each \thanks
% that ends a line with a % and do not let a space in before the next \thanks.
% Spaces after \IEEEmembership other than the last one are OK (and needed) as
% you are supposed to have spaces between the names. For what it is worth,
% this is a minor point as most people would not even notice if the said evil
% space somehow managed to creep in.

% The paper headers
\markboth{Journal of \LaTeX\ Class Files,~Vol.~14, No.~8, August~2015}%
{Shell \MakeLowercase{\textit{et al.}}: Bare Demo of IEEEtran.cls for IEEE Communications Society Journals}
% The only time the second header will appear is for the odd numbered pages
% after the title page when using the twoside option.
% 
% *** Note that you probably will NOT want to include the author's ***
% *** name in the headers of peer review papers.                   ***
% You can use \ifCLASSOPTIONpeerreview for conditional compilation here if
% you desire.

% If you want to put a publisher's ID mark on the page you can do it like
% this:
%\IEEEpubid{0000--0000/00\$00.00~\copyright~2015 IEEE}
% Remember, if you use this you must call \IEEEpubidadjcol in the second
% column for its text to clear the IEEEpubid mark.

% use for special paper notices
%\IEEEspecialpapernotice{(Invited Paper)}

% make the title area
\maketitle

% As a general rule, do not put math, special symbols or citations
% in the abstract or keywords.
\begin{abstract}
In general, the zero-forcing (ZF) precoding suffers from a severe receive signal-to-noise ratio (SNR) degradation in the high interference regime. However, recent evidences from realistic measurements demonstrated that millimeter wave (mmWave) systems are mainly noise-limited as high gain antennas behave as spatial filters to the interference signal. This makes ZF precoding equally attractive as that of other linear precoding counterparts. Considering ZF precoding, this paper aims to derive a lower bound on the sum-rate achieved by a decremental beam selection (BS) algorithm in a beamspace MIMO (B-MIMO) system operating at mmWave frequencies. This bound relates Frobenious norms of precoding matrices of full and reduced dimensional (i.e. after BS) B-MIMO systems through a deterministic square-hyperbolic function. Note that, both ZF precoding and decremental BS are not new concepts. However, the derived sum-rate bound provides a new insight to the topic. Given a particular full dimensional B-MIMO channel, the presented bound can be used to understand limits of BS algorithms.   
\end{abstract}

% Note that keywords are not normally used for peerreview papers.

% For peer review papers, you can put extra information on the cover
% page as needed:
% \ifCLASSOPTIONpeerreview
% \begin{center} \bfseries EDICS Category: 3-BBND \end{center}
% \fi
%
% For peerreview papers, this IEEEtran command inserts a page break and
% creates the second title. It will be ignored for other modes.
\IEEEpeerreviewmaketitle

\section{Introduction}
% The very first letter is a 2 line initial drop letter followed
% by the rest of the first word in caps.
% 
% form to use if the first word consists of a single letter:
% \IEEEPARstart{A}{demo} file is ....
% 
% form to use if you need the single drop letter followed by
% normal text (unknown if ever used by the IEEE):
% \IEEEPARstart{A}{}demo file is ....
% 
% Some journals put the first two words in caps:
% \IEEEPARstart{T}{his demo} file is ....
% 
% Here we have the typical use of a "T" for an initial drop letter
% and "HIS" in caps to complete the first word.
%\iffalse
\IEEEPARstart{M}{}illimeter wave (mmWave) systems have recently been realized as promising candidates for next generations of the wireless technologies. MmWave systems are high beamforming gain MIMO systems, as a large number of antenna elements can be packed in a small aperture area. However, this poses a serious restriction on the implementation of fully digital and optimal beamforming which requires one radio frequency (RF) chain per antenna and results in a high power consumption and hardware cost. Alternatively, sub-optimal approaches proposed in the literature consider either fully digital precoding with low resolution digital-to-analog converters (DACs)~\cite{Roth18} or hybrid digital/analog precoding~\cite{Ayach14} with a reduced number of RF chains and high resolution DACs. Both approaches have their associated advantages and disadvantages as discussed in~\cite{Roth18}. 
\par The B-MIMO~\cite{Brady13} is a recent hybrid beamforming concept which reduces the RF complexity dramatically without any significant performance loss. In contrast to a complex phase shifter/combiner network, the analog beamforming in a B-MIMO system is done by a simple discrete lense antenna array (DLA). The DLA acts as a discrete Fourier transform (DFT) filter and it transforms the spatial multi-user MIMO (MU-MIMO) channel into a sparse beamspace channel. Each beam in a B-MIMO system is connected to a single RF chain. For the RF complexity reduction, B-MIMO system employ a digital precoder to the reduced dimensional beamspace channel followed by a BS module~\cite[see Fig. 4]{Brady13}. Therefore, in order to maximize the sum-rate, BS studies have recently gained considerable attention~\cite{Syed13,Amad15,GAO16,Rahul18}. In general, the BS problem is similar to the antenna/user selection problem which have been extensively studied in literature. In~\cite{Syed13}, BS with more than one beam allocated per user have been studied for different linear precoding schemes. Results show that ZF precoding is spectrally inefficient in high interference scenarios. However, realistic measurement based studies in~\cite{Intr-Iqbal17} and references therein reveal that due to high gain antenna systems, mmWave networks are not interference limited. Amadori et al.~\cite{Amad15}, investigated concepts from antenna selection literature and proposed incremental and decremental BS algorithms based on the ZF precoding. Different from~\cite{Syed13}, algorithms proposed in~\cite{Amad15,GAO16,Rahul18} select one beam per user and provide close approximation to results obtained from a full dimensional B-MIMO system. 
\par Despite a considerable work on BS algorithm development, theoretical studies on performance bounds merely exist. An upper bound introduced in~\cite{Syed13} is based on a non-realistic assumption of perfectly orthogonal channels. As the B-MIMO channels at mmWave frequencies are expected to be sparse; therefore, the upper bounds derived in the antenna selection literature e.g.~\cite{Mol05,Love08} which are based on the Rayleigh fading assumption cannot be extended to beamspace channels. Note that the bounds based on the Rayleigh/Rice fading assumption rely on the fact that inphase and quadrature components in these channels are mutually independent random processes offering two degrees of freedom. However, this is not a case in the sparse multipath channels~\cite{PR13}. This leads us to the primary motivation of this work. The foundation of this work is laid by the paper of Hoog and Mattheij~\cite{hoog2007}, which provides an extensive mathematical treatment on the maximum volume subset selection from \textit{real} matrices. We intend to derive a lower bound on the sum-rate achieved by a ZF-based decremental BS algorithm without any particular assumption on the fading envelope of the channel. 
\iffalse Objective of this bound is to deterministically answer the following: How much the results of a reduced and a high-dimensional full RF complexity system can differ from each other? \fi    
\section{System And Channel Model}
We consider downlink communication from an access point (AP) equipped with a DLA having a maximum of $n_\text{B}$ signal space dimensions. Theoretically, a DLA can be modeled with a critically spaced uniform linear array (ULA) i.e. $n_\text{B}=\frac{2L_a}{\lambda}$, where $L_a$ and $\lambda$ denote the physical length of the ULA and wavelength of the carrier frequency~\cite{Brady13}, respectively. At a particular time instant, the AP can communicate with $n_\text{U}\leq n_\text{B}$ user terminals; each equipped with an omni-directional antenna. The antenna domain multi-user MIMO (MU-MIMO) system with a ULA at the AP is defined as
\begin{equation} 
\label{Sys_model_MIMO}
\tilde{\bm {y}}=\tilde{\bm{H}} \tilde{\bm{F}}\bm {x}+\tilde{\bm {w}}, \tilde{\bm{H}}=\left[
\begin{array}{ c c c c }
\tilde{\bm{h}}_1 & \tilde{\bm{h}}_2 & \cdots & \tilde{\bm{h}}_{n_\text{U}} \\    
\end{array} \right]^T\in \mathbb{C}^{{n_\text{U}}\times n_\text{B}}
\end{equation}
where, $\tilde{\bm{H}}$ is a MU-MIMO channel matrix, $\tilde{\bm{F}}\in \mathbb{C}^{n_\text{B}\times{n_\text{U}}}$ is a transmit precoding matrix and $\bm {x}\in \mathbb{C}^{n_\text{U}\times 1}$ is a transmit signal vector with $E[\bm {x}\bm {x}^{H }]=\bm {I}_{n_\text{U}}$. Let $P$ be the total transmit power, then $\bm {x}$ satisfies the average power constraint $E\left[ \left\| \tilde{\bm{F}}\bm {x}\right\|^{2}\right]  \leq P$. Finally, $\tilde{\bm{w}}$ in~(\ref{Sys_model_MIMO}) is a additive white Gaussian noise vector with $\tilde{\bm {w}}\sim \mathcal {CN}(0,\sigma ^{2}\bm {I}_{n_\text{U}})$. Assuming a propagation channel with a Line-Of-Sight (LOS) path followed by $L$ reflected non-LOS multipath components (MPCs), a generic narrow-band channel model for the $k^\text{th}$ user is given as

\begin{equation} \label{Gchan} 
\tilde{\bm{h}}_{k}=\underbrace{\beta ^{(0)}_{k}\bm {a}\left ({\theta ^{(0)}_{k}}\right)}_\text{LOS path}+ \underbrace{\sum \limits _{\ell =1}^{L}\beta ^{(\ell)}_{k}\bm {a}\left ({\theta ^{(\ell)}_{k}}\right)}_\text{NLOS MPCs}\!\cdot \end{equation}
In~(\ref{Gchan}), $\beta ^{(\cdot)}_{k}$ corresponds to MPC gains. For the ULA, the array steering vector $\bm {a}\left ({\theta ^{(\cdot)}_{k}}\right)\in \mathbb{C}^{n_\text{B}\times 1}$ is defined as $\bm {a}\left ({\theta ^{(\cdot)}_{k}}\right)=\frac{1}{\sqrt{n_\text{B}}}\left [{\exp \left ({-j 2\pi\left( \theta ^{(\cdot)}_{k}\right) i }\right)}\right]_{i\in \mathcal {I}(n_\text{B})}$
\iffalse
\begin{equation} \bm {a}\left ({\theta ^{(\cdot)}_{k}}\right)=\frac{1}{\sqrt{n_\text{B}}}\left [{\exp \left ({-j 2\pi\left( \theta ^{(\cdot)}_{k}\right) i }\right)}\right]_{i\in \mathcal {I}(n_\text{B})}\end{equation} \fi
where, ${\theta ^{(\cdot)}_{k}}=0.5\sin\left(\phi ^{(\cdot)}_{k} \right) $ is the spatial frequency induced by the physical angle $\phi ^{(\cdot)}_{k}\in\left[\frac{-\pi}{2},\frac{\pi}{2} \right]  $and $\mathcal {I}(n_\text{B})=\{i-(n_\text{B}-1)/2, i=0,1,\ldots,n_\text{B}-1\} $ is a symmetric set of indices centered around zero. Since, DLA acts as a DFT filter, the spatial MU-MIMO channel $\tilde{\bm{H}}$ can be transformed into the beamspace channel $\bm {H}$ by multiplying it with a unitary DFT matrix $\bm {U}=\frac {1}{\sqrt {n_\text{B}}} \bm {a}\left ({\theta _{i}=\frac {i}{n_\text{B}}}\right)_{i\in \mathcal {I}(n_\text{B})}\quad \in \mathbb{C}^{n_\text{B}\times n_\text{B}}$,  
\iffalse 
\begin{equation} \bm {U}=\frac {1}{\sqrt {n_\text{B}}} \bm {a}\left ({\theta _{i}=\frac {i}{n_\text{B}}}\right)_{i\in \mathcal {I}(n_\text{B})}\quad \in \mathbb{C}^{n_\text{B}\times n_\text{B}}\end{equation} \fi
where, the columns $\bm{u}_i\in\bm{U}$ are the analog orthogonal precoding vectors corresponding to $n_\text{B}$ spatial modes pointed towards fixed predefined directions.~An equivalent beamspace representation of~(\ref{Sys_model_MIMO}) is given by  
\begin{equation} 
\label{Sys_model_B-MIMO}
\bm {y}=\bm{H} \bm{F}\bm {x}+\bm {w}, \quad   \bm{H}=\tilde{\bm {H}}\bm{U}, \quad   \bm{F}=\bm{U}^H\tilde{\bm {F}}
\end{equation}
where, $\bm{H}\in \mathbb{C}^{{n_\text{U}}\times n_\text{B}}$ is a full dimensional (i.e. without any BS) B-MIMO channel. In this paper, we assume that $\bm{H}$ is known at the AP.
\section{Beam Selection}
Considering ZF precoding $\bm{F}=\bm{H}^+$, where $\left( \cdot\right) ^+$ denotes the pseudo-inverse of the matrix $\bm{H}$, then the achievable sum-rate of a full-dimensional B-MIMO system with $n_\text{B}$ RF chains is described as~\cite{Amad15,GAO16} 
\begin{equation}
\label{SH_rate}
R_\text{Full}=n_\text{U}\log_2\left(1+\frac{P}{\sigma^2 \left\|\bm{H}^+ \right\|_F^2 } \right),
\end{equation} 
where, $\left\|  \bm{H}^+ \right\|^2_F=\Tr\left\lbrace \left( \bm{H}\bm{H}^H\right)^{-1} \right\rbrace $ is Frobenious norm of the pseudo-inverse of $\bm{H}$. In contrast to the full RF complexity B-MIMO system (i.e. with $n_\text{B}$ RF chains), we assume that AP is equipped with $K$ RF chains allowing a selection of $n_\text{U}\leq K\leq n_\text{B}$ beams. Let us decompose the full complexity B-MIMO channel matrix $\bm{H}$ into a reduced dimensional beamspace channel $\bm{H}_s\in \mathbb{C}^{{n_\text{U}}\times K }$ of the selected beams and a set of discarded beams $\bm{H}_d\in \mathbb{C}^{{n_\text{U}}\times\left( {n_\text{B}-K}\right) }$ as
\begin{equation}
\label{Hd}
\bm{H}\bm{\Pi}=\left[
\begin{array}{ c c  }
\bm{H}_s & \bm{H}_d  \\    
\end{array} \right]
\end{equation}
where, $\bm{\Pi}$ is a permutation matrix. Let $\bm{F}_s=\bm{H}^+_s$ be the ZF precoding matrix of the reduced-dimensional B-MIMO system, then sum-rate $R_s\leq R_\text{Full}$ with $K$ selected beams is defined as
\begin{equation}
\label{SHs_rate}
R_s=n_\text{U}\log_2\left(1+\frac{P}{\sigma^2 \left\|\bm{H}^+_s \right\|_F^2 } \right) \cdot
\end{equation} 
From~(\ref{SHs_rate}), it is clear that for sum-rate maximization, a BS algorithm is supposed to minimize $\left\|\bm{H}^+_s \right\|_F^2$. An optimal solution to this problem is computationally prohibitive as it requires exhaustive pseudo-inverse evaluations over $\binom{n_\text{B}}{K}$ combinations. 
\subsection{Decremental Beam Selection Algorithm}
\label{dec_algo}
A greedy algorithm presented here intends to minimize $\left\|  \bm{H}_s^+ \right\|^2_F$ by successive elimination of beams from the full dimensional B-MIMO channel matrix $\bm{H}$. Using $\rank-1$ update of the matrix inverse, a beam index $j_1$ is dropped such that  
\begin{equation}
\label{Dec1}
j_1=\arg\left\lbrace \min_{j=1\cdots n_\text{B} } \Tr \left(  \left( \bm{HH}^H-\bm{h}_j\bm{h}^H_j  \right)^{-1}\right) \right \rbrace \cdotp
\end{equation}
Let $\bm{H}_1\in \mathbb{C}^{{n_\text{U}}\times\left( {n_\text{B}-1}\right)} $ be a B-MIMO channel matrix obtained after dropping a beam $j_1$, then in the second iteration, a beam $j_2$ is dropped such that 
\begin{equation}
\label{Dec2}
j_2=\arg\left\lbrace \min_{j=1\cdots n_\text{B}-1 } \Tr \left(  \left( \bm{H}_1\bm{H}_1^H-\bm{h}_j\bm{h}^H_j  \right)^{-1}\right) \right \rbrace 
\end{equation} 
and so on till $K$ beams are left. A low complexity implementation of the above decremental algorithm can be done using Sherman-Morrison formula for $\rank-1$ update which requires $\mathcal{O}\left(n_\text{B}n_\text{U}^3 \left(n_\text{B}-K \right)  \right) $ operations~\cite{Hager1989}. Under the condition $n_\text{B}\gg n_\text{U}$, this operation count is much lower than the decremental algorithm proposed in~\cite{Amad15} which requires $\mathcal{O}\left(n_\text{B}^3  \right) $ operations. The algorithm in~(\ref{Dec1}) and~(\ref{Dec2}) results in the following bound on the Frobenious norm of $\bm{H}^+_s$.

\begin{theorem}
	\label{theorm1}
	\textit{For $\bm{H}\in \mathbb{C}^{n_\text{U}\times n_\text{B}}$ with $n_\text{U}\leq n_\text{B}$, there exist a permutation matrix $\bm{\Pi}$ such that~(\ref{Hd}) holds for $K\geq n_\text{U}$} with  
	\begin{equation}
	\label{Thrm1}
	\left\|  \bm{H}_s^+ \right\|^2_F \leq \frac{\left( n_\text{B}-n_\text{U}+ 1\right) }{\left( K-n_\text{U}+ 1\right) }  \left\| \bm{H}^+\right\|^2_F
	\end{equation}

\end{theorem}

\textit{Proof: }Let $\bm{h}_j$ be a beamspace (column) channel vector to be eliminated from $\bm{H}$ at the first iteration, i.e., $K=n_\text{B}-1$. Then from~(\ref{Dec1}), the $\Tr \left(\bm{HH}^H-\bm{h}_j\bm{h}^H_j  \right)^{-1}$ would be lowest as compared to any other column $\bm{h}_m\neq\bm{h}_j$ in $\bm{H}$. Therefore, 
\begin{align*}
\Tr \left(\bm{HH}^H-\bm{h}_j\bm{h}^H_j  \right)^{-1}   & \leq \Tr \left(\bm{HH}^H-\bm{h}_m\bm{h}^H_m  \right)^{-1}    \\
&\leq \sum^{n_\text{B}}_{m=1} \Tr \left(\bm{HH}^H-\bm{h}_m\bm{h}^H_m  \right)^{-1} \cdot \numberthis \label{T3-9}
\end{align*}
Now we evaluate the right-hand side (R.H.S.) of~(\ref{T3-9}), which after re-arranging becomes
\begin{equation}
\label{T3-2}
\begin{array} {lcl}
\bm{HH}^H - \bm{h}_m\bm{h}^H_m\\=\bm{HH}^H - \bm{h}_m\bm{h}^H_m \left( \left( \bm{HH}^H\right) ^{-{\frac{1}{2}}}\right)^H \left( \bm{HH}^H\right) ^{\frac{1}{2}} \\ = \left( \left( \bm{HH}^H\right) ^{\frac{1}{2}} - \bm{h}_m\bm{h}^H_m \left( \left( \bm{HH}^H\right) ^{-{\frac{1}{2}}}\right)^H  \right)\left( \bm{HH}^H\right) ^{\frac{1}{2}}  \\ =\left( \bm{HH}^H\right) ^{\frac{1}{2}}\bm{Q}\left( \bm{HH}^H\right) ^{\frac{1}{2}}
\end{array}
\end{equation}
where,
\begin{equation}
\label{T3-3}
\begin{array} {lcl}
\bm{Q}&=&\bm{I} - \left( \bm{HH}^H\right) ^{-{\frac{1}{2}}}\bm{h}_m\bm{h}^H_m  \left( \left( \bm{HH}^H\right) ^{-{\frac{1}{2}}}\right)^H \\&=& \bm{I} - \bm{q}_m \bm{q}^H_m,\quad \text{and}\quad \bm{q}_m=\left( \bm{HH}^H\right) ^{-{\frac{1}{2}}}\bm{h}_m \cdot
\end{array}
\end{equation}
Now the inverse of $\rank-1$ updated matrix in~(\ref{T3-2}) becomes
\begin{equation} 
\label{T3-4}
\left(  \bm{HH}^H - \bm{h}_m\bm{h}^H_m \right)^{-1} = \left( \bm{HH}^H\right) ^{-\frac{1}{2}}\bm{Q}^{-1}\left( \bm{HH}^H\right) ^{-\frac{1}{2}}\cdot
\end{equation}
Using the Sherman-Morrison formula~\cite{Hager1989} for $\rank-1$ update,  $\bm{Q}^{-1}$ can be written as 
\begin{equation}
\label{T3-5}
\begin{array} {lcl}
\bm{Q}^{-1}=\left( \bm{I} - \bm{q}_m \bm{q}^H_m\right) ^{-1} = \bm{I} + \frac{\bm{q}_m \bm{q}^H_m}{1-\bm{q}^H_m\bm{q}_m }= \bm{I }+ \frac{\bm{q}_m \bm{q}^H_m}{1-\left\| \bm{q}_m \right\| ^2_2 }
\end{array}
\end{equation}
Plugging~(\ref{T3-5}) in~(\ref{T3-4}) and substituting the scalar $\tilde{q}= 1-\left\| \bm{q}_m \right\|^2_2 $ for simplification purpose, one can write~(\ref{T3-4}) as
\begin{equation}
\label{T3-6}
\begin{array} {lcl}
\left(  \bm{HH}^H - \bm{h}_m\bm{h}^H_m \right)^{-1}\\=\left( \bm{HH}^H\right) ^{-\frac{1}{2}} \left(  \bm{ I }+ \frac{\bm{q}_m \bm{q}^H_m}{\tilde{q} } \right)     \left( \bm{HH}^H\right) ^{-\frac{1}{2}} \\= \frac{1}{\tilde{q}}\left( \bm{HH}^H\right) ^{-\frac{1}{2}}\left( \tilde{q}\bm{I} + \bm{q}_m \bm{q}^H_m \right) \left( \bm{HH}^H\right) ^{-\frac{1}{2}}\\ =\frac{1}{\tilde{q}} \left( \tilde{q}\left( \bm{HH}^H\right) ^{-1} +\left( \bm{HH}^H\right) ^{-\frac{1}{2}}(\bm{q}_m\bm{q}^H_m)\left( \bm{HH}^H\right) ^{-\frac{1}{2}} \right)\\ =\frac{1}{\tilde{q}} \left( \tilde{q}(\bm{HH}^H)^{-1} +\left( \bm{HH}^H\right) ^{-1}(\bm{h}_m\bm{h}^H_m)\left( \bm{HH}^H\right) ^{-1} \right)
\end{array}
\end{equation}
Substituting $\tilde{q}= 1-\left\| \bm{q}_m \right\|^2_2 $ back in~(\ref{T3-6}), and applying the trace operator followed by summation as on the R.H.S. of~(\ref{T3-9}), we get 
\begin{equation}
\label{Semi_final}
\begin{multlined}
\sum^{n_\text{B}}_{m=1}\left( 1-\left\|  \bm{q}_m\right\| ^2\right) \Tr\left( \bm{HH}^H-\bm{h}_m\bm{h}^H_m\right) ^{-1} \\ = \sum^{n_\text{B}}_{m=1}\left( 1-\left\|  \bm{q}_m\right\| ^2\right)  \Tr \left( \bm{HH}^H\right) ^{-1}\\ \qquad+ \sum^{n_\text{B}}_{m=1}\left[ \Tr \left( \bm{HH}^H\right) ^{-1}\bm{h}_m\bm{h}^H_m \left( \bm{HH}^H\right) ^{-1}\right]                 
\end{multlined}
\end{equation}
Note that,

\begin{equation} 
\label{T3-12}
\begin{split}
\sum^{n_\text{B}}_{m=1}\left\| \bm{q}_m \right\| ^2   & = \Tr\left( \sum_{m=1}^{n_\text{B}}\bm{q}_m\bm{q}^H_m\right)  \\
&=\Tr\left( \left( \bm{HH}^H\right) ^{-\frac{1}{2}}\sum_{m=1}^{n_\text{B}} \bm{h}_m\bm{h}^H_m\left( \bm{HH}^H\right) ^{-\frac{1}{2}}\right) \\
&= \Tr\left( \left( \bm{HH}^H\right) ^{-\frac{1}{2}} \bm{HH}^H\left( \bm{HH}^H\right) ^{-\frac{1}{2}}\right) \\
&= n_\text{U} 
\end{split}
\end{equation}
and  
\begin{figure}[t]
	\includegraphics[width=0.9\columnwidth]{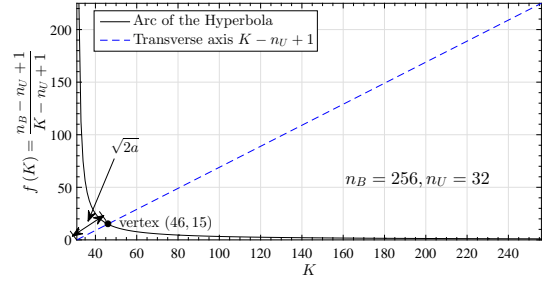}
	\caption{Sharpness analysis of the bound in~(\ref{Thrm1}) as function of $K$}
	\label{mHXPD0}
\end{figure}   
\begin{equation}
\label{T3-13}
\begin{array} {lcl}
\sum_{m=1}^{n_\text{B}}\Tr\left( \left( \bm{HH}^H\right) ^{-1}\bm{h}_m\bm{h}^H_m\left( \bm{HH}^H\right) ^{-1} \right) \\=  \Tr\left( \left( \bm{HH}^H\right) ^{-1}\sum_{m=1}^{n_\text{B}} \bm{h}_m\bm{h}^H_m\left( \bm{HH}^H\right) ^{-1}\right) \\=    
\Tr\left( \left( \bm{HH}^H\right) ^{-1} \bm{HH}^H\left( \bm{HH}^H\right) ^{-1}\right) \\= \left\|  \bm{H}^+ \right\| ^2_F
\end{array}\cdot
\end{equation} 
Therefore~(\ref{Semi_final}) simplifies to
\begin{equation}
\label{Semi_final0}
\sum^{n_\text{B}}_{m=1} \Tr\left( \bm{HH}^H-\bm{h}_m\bm{h}^H_m\right) ^{-1}  = \frac{\left( n_\text{B} - n_\text{U} + 1\right)}{\left( n_\text{B} - n_\text{U} \right)}  \left\| \bm{H}^+\right\| ^2_F               \cdot
\end{equation}
Plugging~(\ref{Semi_final0}) in~(\ref{T3-9}) shows that~(\ref{Thrm1}) holds for $ K=  n_\text{B}-1 $. Proceeding in the same way and deleting columns at each iteration, the result follows by induction.$\Box$%  lamma 3 proof 
\par Note that, due to summation on R.H.S. of~(\ref{T3-9}), the bound in~(\ref{Thrm1}) is likely to be loose when $n_\text{B}\gg K$. Since, the mmWave systems are supposed to operate with large dimensional antenna arrays, and we know from~\cite{GAO16} that a selection of $K=n_\text{U}$ beams (i.e., one beam per user) results in a considerable sum-rate loss as compared to the full RF complexity system, particularly when $n_\text{B}\gg n_\text{U}$. Therefore, in order to obtain a comparable performance as that of full RF complexity system, study of increased RF complexity system (i.e., $K> n_\text{U}\rightarrow$ more than one beam per user) is of fundamental importance. Interestingly, the bound in~(\ref{Thrm1}) demonstrates that both $\left\|  \bm{H}_s^+ \right\|^2_F$ and $\left\|  \bm{H}^+ \right\|^2_F$ are related by a square-hyperbolic function $y=\frac{a}{x}$\footnote{Note that, the assumption $n_\text{B}\gg n_\text{U}$ results in a positive value of $a$. Therefore, resulting square-hyperbola will lie in I and III quadrants having co-ordinate axis as its asymptotes. The arc of hyperbola in the I quadrant corresponds to the case when $K\geq n_\text{U}$ and its arc in the III quadrant corresponds to the case when $K< n_\text{U}$. Since the algorithm is valid only for $K\geq n_\text{U}$, therefore, presented results and discussions in this paper have been limited to analysis of first quadrant as in~Fig.(\ref{mHXPD0}).}, where $a=n_\text{B}-n_\text{U}+ 1$ and $x=K-n_\text{U}+ 1,~ \forall K\geq n_\text{U}$ with center located at the point $\left(n_\text{U}-1 ,0\right) $.~Fig.(\ref{mHXPD0}) shows that for $K> n_\text{U}$, the lower bound in~(\ref{Thrm1}) hyperbolically becomes tighter as $K$ approaches $n_\text{B}$. Notice that, the vertex of the square-hyperbola in~Fig.(\ref{mHXPD0}) lies at the point $\left(n_\text{U}-1+\sqrt{a},\sqrt{a} \right) $. This shows that, a small increment of $\sqrt{a}$ in $K = n_\text{U}-1$ beams reduces maximum euclidean distance between $\left\|  \bm{H}_s^+ \right\|^2_F$ and $\left\|  \bm{H}^+ \right\|^2_F$ from $a$ to $\sqrt{a}$ which demonstrates a very fast convergence with an increase in $K$. 
\subsection{Improving sharpness of~(\ref{Thrm1})}
Assuming that $n_\text{U}$ and $K$ are constants, then sharpness of~(\ref{Thrm1}) can be further improved by reducing $n_\text{B}$. This can be done by pre-selecting a set of candidate beams $n_\text{c}\leq n_\text{B}$ prior to the start of decremental BS. For this, we employ subspace sampling technique introduced in~\cite{Mahoney697} to determine influence score of each beam to be selected as $\pi_i=\frac{\left\| v_{i}\right\|_2^2 }{n_\text{U}} $, where, $v_{i}$ is the $i^\text{th}$ right singular vector of $\bm{H}$. Finally, the submatrix $\bm{H}_c \in \mathbb{C}^{n_\text{U}\times n_\text{c}} $ of $\bm{H}$ with $n_\text{c}$ candidate beams is selected such that the probability of each selected beam satisfies the criterion $p_i=\min\left ( 1,n_\text{B}\pi_i\right )$. Notice that, the subspace sampling technique above selects $n_c$ beams while assuring that $ \left\| \bm{H}_c^+\right\|^2_F= \left( 1-\epsilon\right) \left\| \bm{H}^+\right\|^2_F$, where $\epsilon$ is a small number. Therefore,~(\ref{Thrm1}) becomes
\begin{equation} 
\label{Improv_bound}
\left\|  \bm{H}_s^+ \right\|^2_F\leq\frac{\left( n_\text{c}-n_\text{U}+ 1\right) }{\left( K-n_\text{U}+ 1\right) }  \left\| \bm{H}_c^+\right\|^2_F ,
\end{equation}
where, $\left\| \bm{H}_c^+\right\|^2_F\approx\left\| \bm{H}^+\right\|^2_F$ and $n_\text{c}\ll n_\text{B} $ particularly when $\bm{H}$ is ill-conditioned. Therefore,~(\ref{Improv_bound}) is expected to be much tighter than~(\ref{Thrm1}).~Plugging~(\ref{Improv_bound}) in~(\ref{SHs_rate}) gives the following lower bound on the sum-rate $R_s$ as
 
\begin{align*} 
R_s &\geq  n_\text{U}\log_2\left(1+\frac{\left( K - n_\text{U} +1\right)}{\left( n_\text{c} - n_\text{U} + 1\right)} \cdot \frac{P}{\sigma^2 \left\|\bm{H}_\text{c}^+ \right\|_F^2 } \right) \cdot \numberthis \label{SRate_Bound}
\end{align*}
\section{Simulation Results}

\begin{figure}[t]
	\includegraphics[width=1\columnwidth]{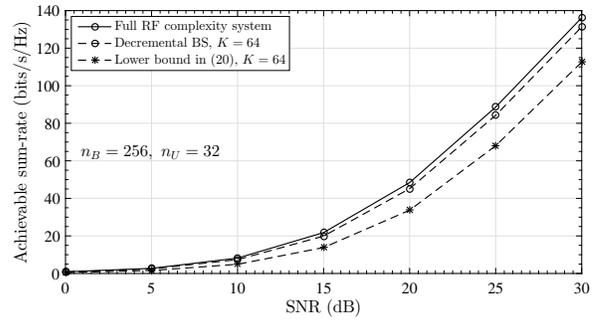}
	\caption{Sum-rate analysis of decremental BS and derived lower bounds.}
	\label{mHXPD1}
\end{figure}

In this section, we analyze a downlink B-MIMO system with a total of $n_\text{B}=256$ beams which are used to serve $n_\text{U}=32$ users. Each user to the AP link is established by a LOS path with a gain described by $\beta ^{(0)}_{k}\sim \mathcal{CN}\left( 0,1\right) $ followed by $L=2$ reflected MPCs having gains $\beta ^{(l)}_{k}\sim \mathcal{CN}\left( 0,10^{-1}\right) $ and $\phi ^{(\cdot)}_{k}$ is a uniformly distributed random variable within $\left[\frac{-\pi}{2},\frac{\pi}{2} \right]$. These assumptions on the channel model are same as the ones used in~\cite{GAO16}. Considering different channel conditions, results in~\cite{Amad15} already provided an extensive comparison of decremental BS with other greedy channel subset selection strategies known from antenna selection literature. Therefore, results in~Fig.~\ref{mHXPD1} are restricted to analysis of derived sum-rate bounds only. For the comparison purpose, we have also shown sum-rate results obtained from~(\ref{Thrm1}).~Fig.~\ref{mHXPD1} shows that a reduced RF complexity system with $K=64$ RF chains provides similar sum-rate guarantees as that of full RF complexity system and the sum-rate bound achieved from~(\ref{Improv_bound}) for $K=64$ is also quite tight. However, for $K=32$, the sum-rate obtained from both~(\ref{Thrm1}) and~(\ref{Improv_bound}) are quite loose, which is due to the summation on R.H.S. in~(\ref{T3-9}).

\section{Conclusion}
\label{sec:CONCLUSION}
A greedy decremental beam selection algorithm with ZF precoding is analytically studied for the B-MIMO systems and a lower sum-rate bound of the algorithm is derived. It has been shown that the maximum euclidean distance between the Frobenious norms of the full and reduced dimensional (i.e. after beam selection) ZF precoding matrices is related by a square-hyperbolic function $\frac{n_\text{B}-n_\text{U}+ 1}{ K-n_\text{U}+ 1 } $. This result shows that the derived bound hyperbolically becomes tighter as $K\geq n_\text{U}$. 

% if have a single appendix:
%\appendix[Proof of the Zonklar Equations]
% or
%\appendix  % for no appendix heading
% do not use \section anymore after \appendix, only \section*
% is possibly needed

% use appendices with more than one appendix
% then use \section to start each appendix
% you must declare a \section before using any
% \subsection or using \label (\appendices by itself
% starts a section numbered zero.)
%

% Can use something like this to put references on a page
% by themselves when using endfloat and the captionsoff option.
\ifCLASSOPTIONcaptionsoff
  \newpage
\fi

% trigger a \newpage just before the given reference
% number - used to balance the columns on the last page
% adjust value as needed - may need to be readjusted if
% the document is modified later
%\IEEEtriggeratref{8}
% The "triggered" command can be changed if desired:
%\IEEEtriggercmd{\enlargethispage{-5in}}

% references section

% can use a bibliography generated by BibTeX as a .bbl file
% BibTeX documentation can be easily obtained at:
% http://mirror.ctan.org/biblio/bibtex/contrib/doc/
% The IEEEtran BibTeX style support page is at:
% http://www.michaelshell.org/tex/ieeetran/bibtex/
%\bibliographystyle{IEEEtran}
% argument is your BibTeX string definitions and bibliography database(s)
%\bibliography{IEEEabrv,../bib/paper}
%
% <OR> manually copy in the resultant .bbl file
% set second argument of \begin to the number of references
% (used to reserve space for the reference number labels box)

%\bibliographystyle{IEEEtran}
%\bibliography{./References_x}

% Generated by IEEEtran.bst, version: 1.14 (2015/08/26)

% that's all folks
\end{document}